**Tunable Coupling between Surface States of a Three-Dimensional Topological Insulator in the Quantum Hall Regime**

Su Kong Chong,[1] Kyu Bum Han,[2] Taylor D. Sparks,[2] and Vikram V. Deshpande[1]*

[1]Department of Physics and Astronomy, University of Utah, Salt Lake City, Utah 84112 USA

[2]Department of Materials Science and Engineering, University of Utah, Salt Lake City, Utah 84112 USA

*Corresponding author: vdesh@physics.utah.edu

The paired top and bottom Dirac surface states, each associated with a half-integer quantum Hall (QH) effect, and a resultant integer QH conductance ($\nu e^2/h$), are hallmarks of a three-dimensional (3D) topological insulator (TI). In a dual-gated system, chemical potentials of the paired surface states are controlled through separate gates. In this work, we establish tunable capacitive coupling between the surface states of a bulk-insulating TI BiSbTeSe$_2$ and study the effect of this coupling on QH plateaus and Landau level (LL) fan diagram via dual-gate control. We observe non-linear QH transitions at low charge density in strongly-coupled surface states, which are related to the charge-density-dependent coupling strength. A splitting of the N= 0 LL at the charge neutrality point for thin devices (but thicker than the 2D limit) indicates inter-surface hybridization possibly beyond single-particle effects. By applying capacitor charging models to the surface states, we explore their chemical potential as a function of charge density and extract the fundamental electronic quantity of LL energy gaps from dual-gated transport and capacitance measurements. These studies are essential for the realization of exotic quantum effects such as topological exciton condensation.

The unique topologically-protected metallic surface states in three-dimensional topological insulators (3D TIs) have been extensively probed experimentally to realize exotic quantum phenomena [1-3]. Time-reversal symmetry guarantees the pairing of top and bottom Dirac surface states in 3D TIs [4,5]. An external perpendicular magnetic field applied can simultaneously break the time-reversal symmetry and manifest in the quantum Hall effect (QHE) in 3D TIs. The half-integer QHE with Landau level (LL) filling factors $v_{t,b}=(N_{t,b}+1/2)$ for top, bottom surfaces is a signature of the Dirac surface states [6-10].

Fine control of chemical potentials of the top and bottom surface states has been central to research in 3D TIs [6-14]. Dual electrostatic gating is an effective method as it provides an additional degree of tuning on both surfaces than a single-gate configuration [15-19]. Independent gate control of the decoupled top and bottom surface states in the quantum Hall regimes has been reported [15,16]. However, the study of the QHE in capacitively-coupled surface states, which serves as the starting point for intriguing quantum states such as topological exciton condensates [20,21], is still lacking. Several groups have demonstrated dual-gate control of the bulk insulating $Bi_{2-x}Sb_xTe_{3-y}Se_y$-based 3D TI in weak and moderate surface-state coupling [17-19]; however, the quality of those devices prevented the observation of QH states.

Here we investigate the effect of capacitive coupling of paired TI surface states on QH plateau development. The thickness-dependent coupling between the top and bottom surfaces is studied by using a dual-gating configuration via a van der Waals (vdW) platform [22]. We study BSTS flake thicknesses down to 10 nm, over a range of inter-surface coupling. We also explore LL formation at different charge propagation regimes of the surface states with capacitive coupling and study their LL energy gaps.

Variable thickness BSTS devices from 89 nm to 10 nm were fabricated (Fig. S1 [23]) for this study. The complete device structure consists of vdW five-layer heterostructures of graphite (Gr)/hexagonal boron nitride (hBN) encapsulated BSTS as inserted in Fig. 1(e). The top and bottom Gr and hBN layers serve as the gate-electrode and dielectric, respectively. This vdW heterostructure is effective in controlling the charge density of the TI surface states [22]. Fig. S1(a)-(d) are the device images, and the device specifications are summarized in Table S1 [23].

Color maps of dual-gated longitudinal resistance ($R_{xx}$) of the respective BSTS devices are shown in Fig. 1(a)-(d). By tuning the top-gate voltage ($V_{tg}$) and bottom-gate voltage ($V_{bg}$), the top and bottom surface states are tuned separately to the two independent ambipolar transport. The red and black dashed lines in the 2D color maps illustrate the Dirac points of the top and bottom surface states, respectively. The two Dirac points intersect and form an $R_{xx}$ maximum at intersection, the overall charge neutrality point (CNP). These two lines divide the $R_{xx}$ map into four quadrants, corresponding to hole-hole (h-h), electron-electron (e-e), hole-electron (h-e), and electron-hole (e-h) conduction of the (top-bottom) surface as labeled in Fig. 1(a).

For 89 nm BSTS (Fig 1(a)), the $R_{xx}$ map shows four nearly-equal-sized conduction quadrants versus the dual-gate voltages. The top and bottom surface states are tuned independently by the top- and bottom-gate, implying negligible capacitive coupling due to the relatively large spatial separation in the bulk. As the thickness of BSTS reduces, the top and bottom Dirac points tend toward the diagonal direction as shown in Fig. 1(b)-(d). The diagonal feature of the $R_{xx}$ maximum is a result of the strong capacitive coupling between the top and bottom surfaces, as observed in similar compounds [18,19].

Another feature arising from the coupling is the overlapping of both surface Dirac points in transport as a function of any one gate voltage. $R_{xx}$ as a function of $V_{bg}$ for 89 nm BSTS taken at three different $V_{tg}$ across the top Dirac point (Fig. 1(e)) show the peak position of $R_{xx}$ (bottom Dirac point) is nearly independent of the density of the top surface. Whereas for 31 nm BSTS, the $R_{xx}$ line profiles (Fig. 1(f)) show a downshift in the bottom Dirac point with increase in $V_{tg}$. In addition, the $R_{xx}$ shows a broadening at both high-density hole and electron conductions, corresponding to the top surface Dirac point as indicated by the arrows in Fig. 1(f). This feature is more apparent in 16 nm BSTS, where the bottom and top Dirac points shift oppositely toward each other, resulting in the broad double $R_{xx}$ peaks in $V_{bg}$ (Fig. 1(g)). The double Dirac points eventually overlap to form a single $R_{xx}$ peak with higher resistance value as shown in 10 nm BSTS (Fig. 1(h)).

Dual-gated magneto-transport at perpendicular magnetic field of 18 T for different thickness BSTS are studied in Fig. 2. The well-developed QH plateaus at high magnetic field give rise to clear QH boundaries in 2D maps of $\sigma_{xx}$ and $\sigma_{xy}$ versus dual-gate voltages. The dashed lines in the maps are tracelines at the boundaries of QH steps in the parameter space of two gate voltages. For 89 nm BSTS (Fig. 2(a) and (b)), the QH boundaries traced with vertical and horizontal straight lines implies that the LLs formed at the top and bottom surface states are completely independent of each other. The line profiles (Fig. 2(c)) reveal equally-spaced QH plateaus in $\sigma_{xy}$ as tuned by $V_{bg}$ into integer increments of $e^2/h$, along with minimum $\sigma_{xx}$ in the QH regimes. The coupling effect between surface states results in the development of a narrower $v=0$ plateau near the overall CNP due to the squeezing of counter-propagating regions (in the 2D map) in thinner BSTS (Fig. 2(d)-(e)). The ($v_t$, $v_b$) indexed in 31 nm (Fig. 2(f)) reveal a sign change in $v_t$ due to the crossing of $N_t=0$ level, as compared with the constant $v_t$ in 89 nm BSTS. The strongly-coupled surface states (≤16 nm) reveal a further squeezing in $v=0$ LL due to the diagonal tendency of both $N_t$ and $N_b=0$ lines (Fig. 2(g)-(l)). The LL fan diagram for any surface transforms from asymmetric into highly symmetric between the hole and electron LLs with $v= 0$ plateau residing at the Dirac point of the surface as shown in Fig. S2 [23].

Referring to the geometry modulation model of QH plateaus (Fig. S3 [23]), the quadrilateral plateaus developed near the overall CNP extend in the diagonal direction and modify into a rhombus shape as the coupling effect becomes appreciable. In the low charge density region, we observe a splitting in both $N_t$ and $N_b= 0$ LLs at the overall CNP as the thickness of BSTS reduces to 16 nm (Fig. 2(g)). The line profile of $\sigma_{xx}$ reveals a dip with zeroth-plateau formed in $\sigma_{xy}$ as presented in Fig. 2(i). Similar feature is observed in 10 nm BSTS (Fig. 2(l)) with more pronounced 0 LL peak splitting. The displacement field (D) versus total charge density (n) maps of $\sigma_{xx}$ and $\sigma_{xy}$ (Fig. S4 [23]) are plotted to extract the total charge density of the 0 LL splitting ($\Delta n$) for the thin BSTS. The $n = n_b + n_t$ is calculated as $n_b = C_{bg}\left(V_{bg} - V_D\right)$ and $n_t = C_{tg}\left(V_{tg} - V_D\right)$ [27]. The $\Delta n$ as a function of BSTS thickness (d) is inserted in Fig. 3(a).

A direct interpretation of the 0 LL splitting is a consequence of the hybridization between top and bottom surface states [28-31]. As the thickness reduces to the thin limit of 3D TI, the inter-surface tunneling due to proximity of the surface states leads to an energy gap at CNP [32]. A key signature of such inter-surface hybridization is the degeneracy lifting of the N= 0 levels in Landau quantization [28-30], which is consistent with our observation. The asymmetry in hole and electron N= 0 sublevels (Fig. 2(l)) in 10 nm BSTS could be explained by the presence of the Zeeman effect [28,29].

However, the inter-surface hybridization in our 16 nm and 10 nm BSTS seems to contradict the well-known 2D limit of 5 nm for $Bi_2Se_3$ [33]. One possibility is the limitation of angle-resolved photoemission spectroscopy in resolving sub-meV energy scale. This is evidenced by the sub-meV hybridization gap signatures in 12-17 nm $Bi_2Se_3$ from phase coherent transport [27]. To check this, we examined the weak anti-localization (WAL) effect by measuring magneto-conductivity ($\Delta\sigma_{xx}$) for the capacitively-coupled BSTS devices at low magnetic field, as shown in Fig. S5 [23]. We observed a consistent transition in WAL [27], indicating a first crossover of 2D limit in 16 nm BSTS, and suppression of WAL by stronger hybridization for even thinner BSTS (8 nm) [34,35]. A sub-meV gap size can easily be obscured by disorder, therefore not visible in our zero magnetic field transport (Fig. 1). The fact that the gap feature shows up in the form of N= 0 LL splitting at strong magnetic field indicates a mechanism of field-dependent hybridization gap as discussed in [31]. Higher magnetic fields will be required for us to check the magnetic-field-proportionality of the gap size. Nonetheless, we cannot exclude the possibility of a many-particle gap developed by the topological exciton condensate as it is also predicted in the same regime [20].

In high-density region, the $N_t$ and $N_b$= 0 tracelines intercross at different angles, resulting in the non-linear QH boundaries as shown in Fig. 2(d) and (e). We assign the non-linearity to the charge density dependent capacitive-coupling between the top and bottom surface states, where screening in the bulk of the sample is weaker at low charge density. This leads to a pronounced bending of $N_t$= 0 and $N_b$= 0 tracelines near the overall CNP as observed in thinner BSTS (16 nm and 10 nm in Fig. 2(g, h) and (j, k)).

The above non-linear QH boundaries features in dual-gated transport are further analyzed to study the capacitive-coupling effect in thin 3D TIs. Fig. 3(a) presents the difference between top surface Dirac point and overall CNP voltages ($V_{tD}$-$V_D$) as a function of $n_b$ extracted from $R_{xx}$ maps for various thickness BSTS at 18 T. The 89 nm and 47 nm BSTS show a nearly linear relation of $V_{tD}$-$V_D$ with $n_b$. In contrast, the non-linear feature manifests in all other (thinner) BSTS. Schematics of linear band of the top surface in Fig. 3(a) illustrates the top surface charge density tuned from the overall CNP by $V_{bg}$. The change in value of $V_{tD}$ is solely controlled by $V_{bg}$ as the chemical potential of the top surface is fixed at its Dirac point. The charge density corresponding to the change of top surface Dirac point from the overall CNP ($n_{tD}$) and then be calculated as $n_{tD} = C_{tg}(V_{tD} - V_D)$.

The $n_{tD}$ as a function of d with $n_b$ fixed at $1\times10^{12}$ cm$^{-2}$ is plotted in Fig. 3(b). The thickness dependence of $n_{tD}$ is studied at high density of $n_b$ to circumvent the non-linear bending effect near the overall CNP. $n_{tD}$ increases monotonically with the reduction in d, which indicates a constant increasing of the electric field penetrating through the bottom surface and the interior bulk layer as the thickness is reduced. The linear extrapolation (black dashed line) of the data points intercepts with the y- and x-axes at $1\times10^{12}$ cm$^{-2}$ and ~60 nm, respectively. The y-intercept at $n_{tD} = n_b$ means the bottom gate tunes an equal amount of density in top and bottom surface states, which is the zero-thickness limit (ignoring the hybridization). The x-intercept suggests the thickness of the BSTS where the surface states are decoupled capacitively, consistent with observations from literature [36,37].

Considering the top and bottom surface states of a thin 3D TI to be a parallel-plate capacitor with an interior bulk insulating layer, together with the top and bottom-gate layers forming a series of three parallel-capacitors, we implement the capacitor charging equations for dual-gated surface states as formulated in [17,18]: $e\Delta n_b = C_{bg}\left(\Delta V_{bg} - \frac{\Delta \mu_b}{e}\right) - C_{BSTS}\left(\frac{\Delta \mu_b}{e} - \frac{\Delta \mu_t}{e}\right)$ − (1), and $e\Delta n_t = C_{tg}\left(\Delta V_{tg} - \frac{\Delta \mu_t}{e}\right) - C_{BSTS}\left(\frac{\Delta \mu_t}{e} - \frac{\Delta \mu_b}{e}\right)$ − (2), where $C_{bg}$, $C_{tg}$ and $C_{BSTS}$ are the top-gate, bottom-gate and BSTS bulk capacitances, and $\Delta\mu_{b(t)}$ is the change in chemical potential of the bottom (top) surface state. These two equations are simplified to two linear relations under the condition where the top surface state stays at its Dirac point, meaning $\Delta n_t$ and $\Delta \mu_t$ have both vanished. The outcome of Eq. (1) is the linear dependence of $\Delta V_{tg}$ versus $\Delta V_{bg}$, and the slope (S) can be expressed as $S = \left(\dfrac{1}{C_{tg}}\right)\left(\dfrac{C_{bg}C_{BSTS}}{C_{bg}+C_{BSTS}}\right)$. The slopes of $\Delta V_{tg}$ versus $\Delta V_{bg}$ in zero magnetic field 2D maps of $R_{xx}$ for different BSTS thickness as shown in the red lines in Fig. 1(b)-(d) are used to estimate the BSTS bulk capacitances. This is to ensure that the slope is not affected by localized states formed in magnetic field. Details of the fitted S, $C_{bg}$, $C_{tg}$ and $C_{BSTS}$ are listed in Table S2 [23]. An average dielectric constant of BSTS ($\varepsilon_{BSTS}$) of about 28 is obtained from different thickness BSTS as shown in Fig. 3(b), which is comparable to similar compounds [17,38].

Chemical potential of the bottom surface state ($\mu_b$) can be evaluated from $\Delta V_{tg}$ by using the relation derived from Eq. (2) $\mu_b = -\dfrac{C_{tg}}{C_{BSTS}} e\Delta V_{tg}$, where the $\Delta V_{tg}$ is the difference between $V_{tD}$ and the overall CNP position ($V_D$). Fig. 3(c) shows the plots of $\mu_b$ as a function of $n_b$ for the 16 nm BSTS at magnetic field of 0T and 18T. The increase in $\mu_b$ in magnetic field is related to the localized electronic states in the bottom surface due to LL formation. The color-shaded regions emphasize density regions in terms of LL indices of bottom surface ($N_b$) between -2 and +2. The change in bottom surface chemical potential with field, $\Delta\mu_b = \mu_b(18T)-\mu_b(0T)$ is inserted in Fig. 3(c). The LL gaps ($\Delta_{\pm 1}$) are estimated from the $\Delta\mu_b$ in the charge density region for $N_b = \pm 1$. Fig. 3(d) displays the $\Delta_{\pm 1}$ as a function of magnetic field for the 16 nm BSTS. The details of the estimated of the $\Delta_{\pm 1}$ at different magnetic fields are given in Fig. S6 [23]. The black dashed curve in Fig. 3(d) serves as a comparison between $\Delta_{\pm 1}$ and the Dirac LL energy relation $\Delta_N = \text{sgn}(N)v_F\sqrt{2e\hbar|N|B}$ [39], with the Fermi velocity ($v_F$) of the BSTS taken to be ~$3\times 10^5$ m/s [6]. Inset in Fig. 3(d) shows nearly linear fitting of $\Delta_{\pm 1}$ with the square root of magnetic field, indicating a good match with the theory despite a ~30% deviation from theoretical values.

Despite the agreement of the $\Delta_{\pm 1}$ obtained from dual-gated transport with the theoretical calculation, the effect of magnetic field on the capacitance of BSTS cannot be ruled out from the study. As the $\mu_b$ function is inversely proportional to $C_{BSTS}$, the reduction in $C_{BSTS}$ can also cause an increment in $\mu_b$. To verify this, we perform dual-gated capacitance measurement on the same device by using a capacitance bridge method [40,41]. A similar diagonal zero-field dual-gated $C_Q$ map (Fig. S7(a) [23]) again verifies the coupling effect of top and bottom surface states. In a magnetic field of 9T (Fig. S7(b) [23]), the $C_Q$ forms dips in the QH regimes corresponding to electronic density of states dips developed in those regimes. Fig. S7(c) compares the line profiles of $C_Q$ as a function of $V_{tg}$ at 0T and 9T at fixed $V_{bg}$. The capacitance values are nearly constant with magnetic field, which indicates an insignificant effect of magnetic field on BSTS bulk capacitance, in agreement with the analysis in [17]. In addition, the LL gap of $\Delta_{\pm 1}$ calculated from $C_Q$ at 9T is about 19 meV, which is close to the value obtained from the dual-gated transport.

In summary, the direct correlation between flake thickness and capacitive coupling gives rise to a tunable coupling of topological surface states. A direct manifestation of the capacitive coupling between surface states is the diagonal feature in dual-gated conduction. In perpendicular magnetic field, the strongly-coupled surface states develop into a series of rhombus-shaped QH plateaus versus dual-gates, and the LL fan diagram transforms into an electron-hole symmetric LL series about the N= 0 LL. A careful analysis reveals non-linear QH boundaries and splitting of N= 0 LL at the overall CNP. We attribute the splitting to a consequence of inter-surface hybridization at high magnetic field. From analysis of non-linear QH boundaries (in this case, $N_t$= 0 LL with $V_{bg}$), we estimated a physical quantity: dielectric constant of BSTS, and an electronic quantity: LL energy separation. The study of surface state coupling effects in QH regimes is believed to pave the path for exotic quantum phenomena such as topological exciton condensation.

**Acknowledgments.** This work was supported by the NSF MRSEC program at the University of Utah under grant # DMR 1121252 and by ACS PRF grant # 58164. A portion of this work was performed at the National High Magnetic Field Laboratory, which is supported by National Science Foundation Cooperative Agreement No. DMR-1644779 and the State of Florida.

**References**

[1] M. Z. Hasan, and C. L. Kane, Rev. Mod. Phys. **82**, 3045 (2010).

[2] M. X. Wang *et al.*, Science **336**, 52 (2012).

[3] M. Mogi *et al.*, Nat. Mater. **16**, 516 (2017).

[4] L. Fu, C. L. Kane, and E. J. Mele, Phys. Rev. Lett. **98**, 106803 (2007).

[5] M. Z. Hasan, and J. E. Moore, Annu. Rev. Condens. Matter Phys. **2**, 55 (2011).

[6] Y. Xu *et al.*, Nat. Phys. **10**, 956 (2014).

[7] R. Yoshimi *et al.*, Nat. Commun. **6**, 6627 (2015).

[8] N. Koirala *et al.*, Nano Lett. **15**, 8245 (2015).

[9] S. Buchenau *et al.*, 2D Mater. **4**, 015044 (2017).

[10] E. J. König, P. M. Ostrovsky, I. V. Protopopov, I. V. Gornyi, I. S. Burmistrov, and A. D. Mirlin, Phys. Rev. B **90**, 165435 (2014).

[11] K. B. Han *et al.*, Sci. Rep. **8**, 17290 (2018).

[12] G. Zhang *et al.*, Adv. Funct. Mater. **21**, 2351 (2011).

[13] D. Kim *et al.*, Nat. Phys. **8**, 459 (2012).

[14] T. Arakane *et al.*, Nat. Commun. **3**, 636 (2012).

[15] Y. Xu, I. Miotkowski, and Y. P. Chen, Nat. Commun. **7**, 11434 (2015).

[16] C. Li *et al.*, Phys. Rev. B **96**, 195427 (2017).

[17] V. Fatemi *et al.*, Phys. Rev. Lett. **113**, 206801 (2014).

[18] F. Yang *et al.*, ACS Nano **9**, 4050 (2015).

[19] A. A. Taskin *et al.*, Nat. Commun. **8**, 1340 (2017).

[20] B. Seradjeh, J. E. Moore, and M. Franz, Phys. Rev. Lett. **103**, 066402 (2009).

[21] M. P. Mink, H. T. C. Stoof, and R. A. Duine, Phys. Rev. Lett. **108**, 186402 (2012).

[22] S. K. Chong *et al.*, Nano Lett. **18**, 8047 (2018).

[23] See Supplemental Material for experimental methods, BSTS device specifications, field-dependent $\sigma_{xy}$ and $\sigma_{xx}$, quantum Hall plateaus model, displacement field and total charge density relation, weak antilocalization analysis, estimation of dielectric constant, field-dependent chemical potential, and quantum capacitance, which includes Refs. [24-26].

[24] Y. Zhang *et al.*, Nature **459**, 820 (2009).

[25] J. Martin, B. E. Feldman, R. T. Weitz, M. T. Allen, and A. Yacoby, Phys. Rev. Lett. **105**, 256806 (2010).

[26] G. L. Yu *et al.*, PNAS **110**, 3282 (2013).

[27] D. Kim, P. Syers, N. P. Butch, J. Paglione, and M. S. Fuhrer, Nat. Commun. **4**, 2040 (2013).

[28] M. Tahir, K. Sabeeh, and U. Schwingenschlögl, J. Appl. Phys. **113**, 043720 (2013).

[29] Z. Yang, and J. H. Han, Phys. Rev. B **83**, 045415 (2011).

[30] A. A. Zyuzin, and A. A. Burkov, Phys. Rev. B **83**, 195413 (2011).

[31] A. Pertsova, C. M. Canali, and A. H. MacDonald, Phys. Rev. B **91**, 075430 (2015).

[32] A. Pertsova, and C. M. Canali, New J. Phys. **16**, 063022 (2014).

[33] Y. Zhang *et al.*, Nat. Phys. **6**, 584 (2010).

[34] M. Lang *et al.*, Nano Lett. **13**, 48 (2013).

[35] A. A. Taskin, S. Sasaki, K. Segawa, and Y. Ando, Phys. Rev. Lett. **109**, 066803 (2012).

[36] B. Skinner, T. Chen, and B. I. Shklovskii, J. Exp. Theor. Phys. **117**, 579 (2013).

[37] T. Bomerich, J. Lux, Q. T. Feng, and A. Rosch, Phys. Rev. B **94**, 075204 (2017).

[38] N. P. Butch *et al.*, Phys. Rev. B **81**, 241301(R) (2010).

[39] J. G. Analytis, R. D. McDonald, S. C. Riggs, J.-H. Chu, G. S. Boebinger, and I. R. Fisher, Nat. Phys. **6**, 960 (2010).

[40] A. F. Young *et al.*, Phys. Rev. B **85**, 235458 (2012).

[41] B. M. Hunt *et al.*, Nat. Commun. **8**, 948 (2017).

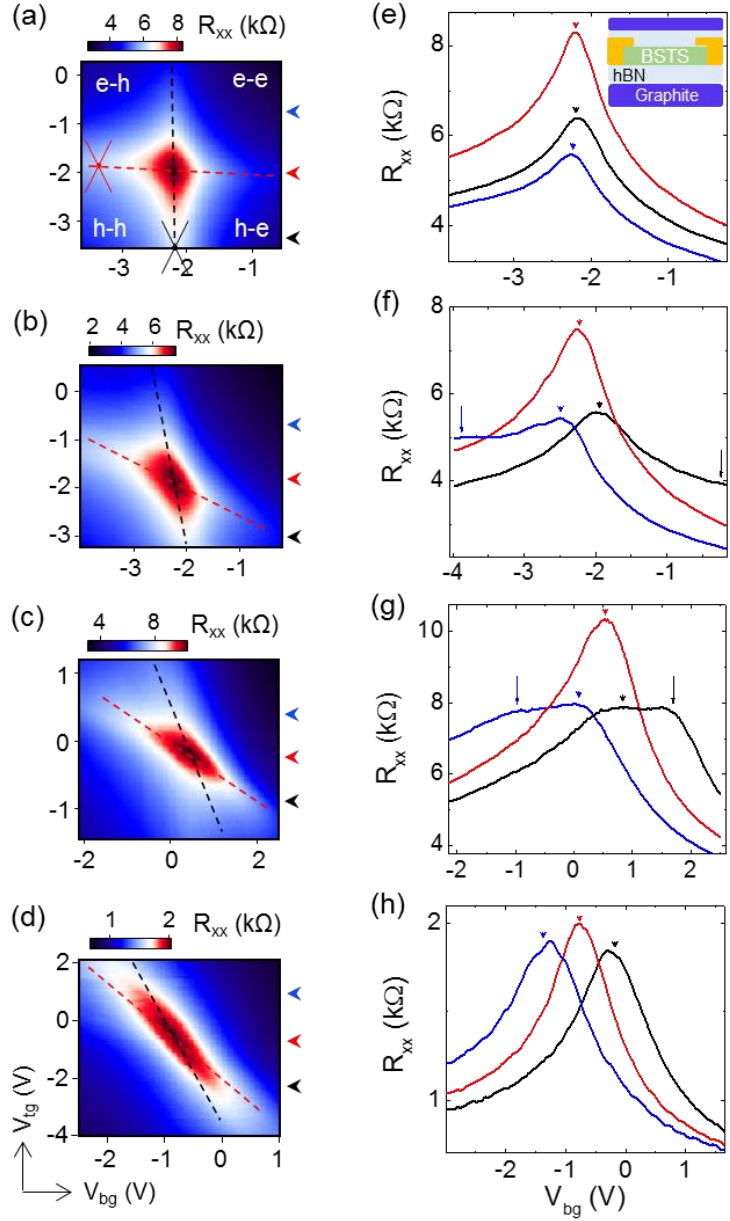

FIG. 1. 2D color maps of $R_{xx}$ as functions of $V_{tg}$ and $V_{bg}$ for BSTS devices with flake thickness of (a) 89 nm, (b) 31 nm, (c) 16 nm, and (d) 10 nm. (e)-(h) Plots of $R_{xx}$ versus $V_{bg}$ extracted from the corresponding maps in (a)-(d) at different $V_{tg}$ indicated by the arrows. Inset in (e) is a schematic of cross-sectional illustration of the device.

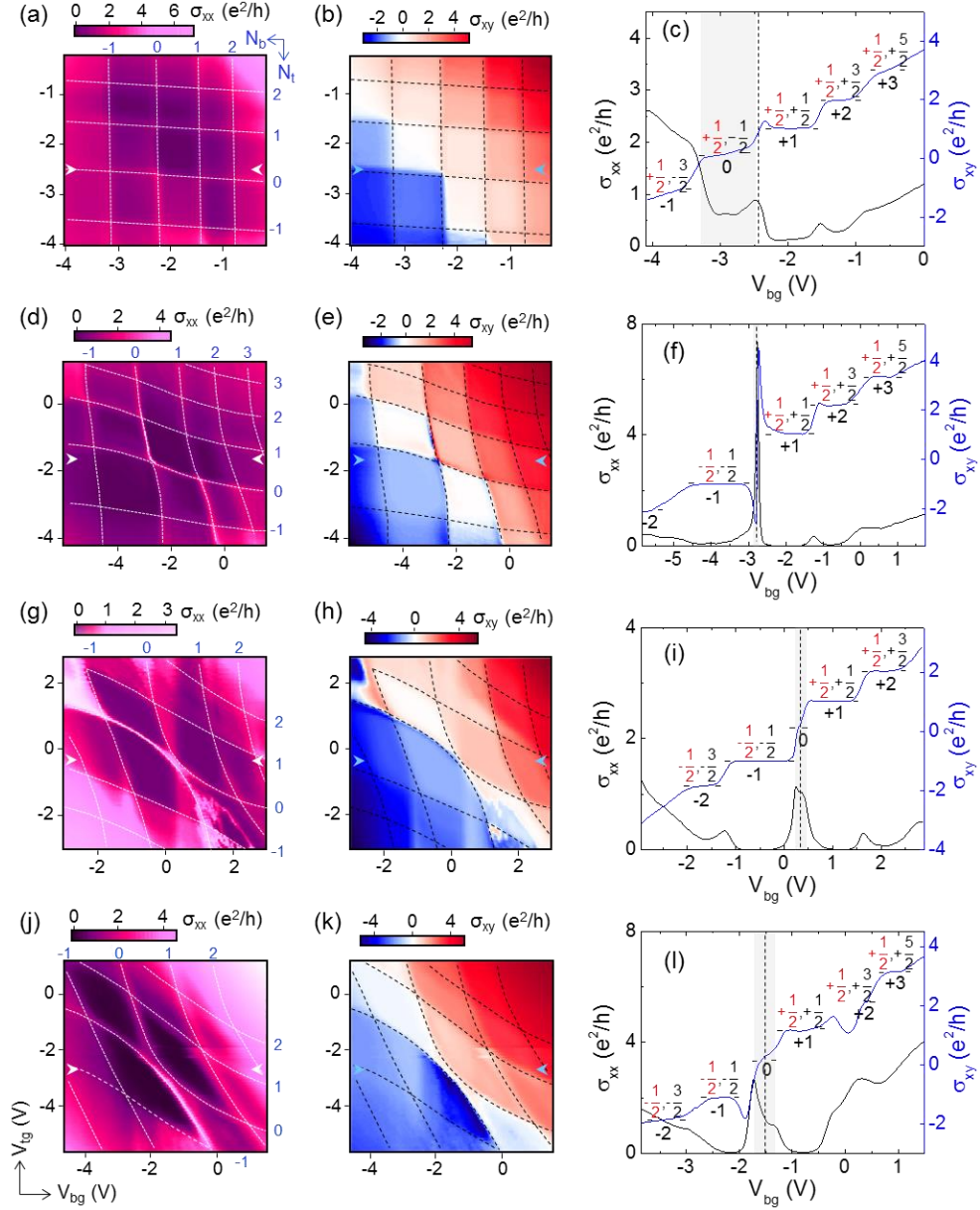

FIG. 2. 2D color maps of $\sigma_{xx}$ and $\sigma_{xy}$ as a function of dual-gated voltages for the (a, b) 89 nm, (d, e) 31 nm, (g, h) 16 nm, and (j, k) 10 nm BSTS devices measured at 18 T. The LL indices for top and bottom surfaces, $N_t$ and $N_b$ are labeled in y-axis and x-axis, respectively. Line profiles of $\sigma_{xx}$ and $\sigma_{xy}$ as a function of $V_{bg}$ extracted from the corresponding maps near the overall CNP (as indicated by the arrows) for the (c) 89 nm, (f) 31 nm, (i) 16 nm, and (l) 10 nm BSTS devices. The vertical black dashed line and grey highlight in (c), (f), (i), and (l) denote the overall CNP and $v=$ 0 QH plateau, respectively. The LL filling factors for top and bottom surfaces ($v_t$, $v_b$) and total, $v = v_t + v_b$ are indexed in the figure.

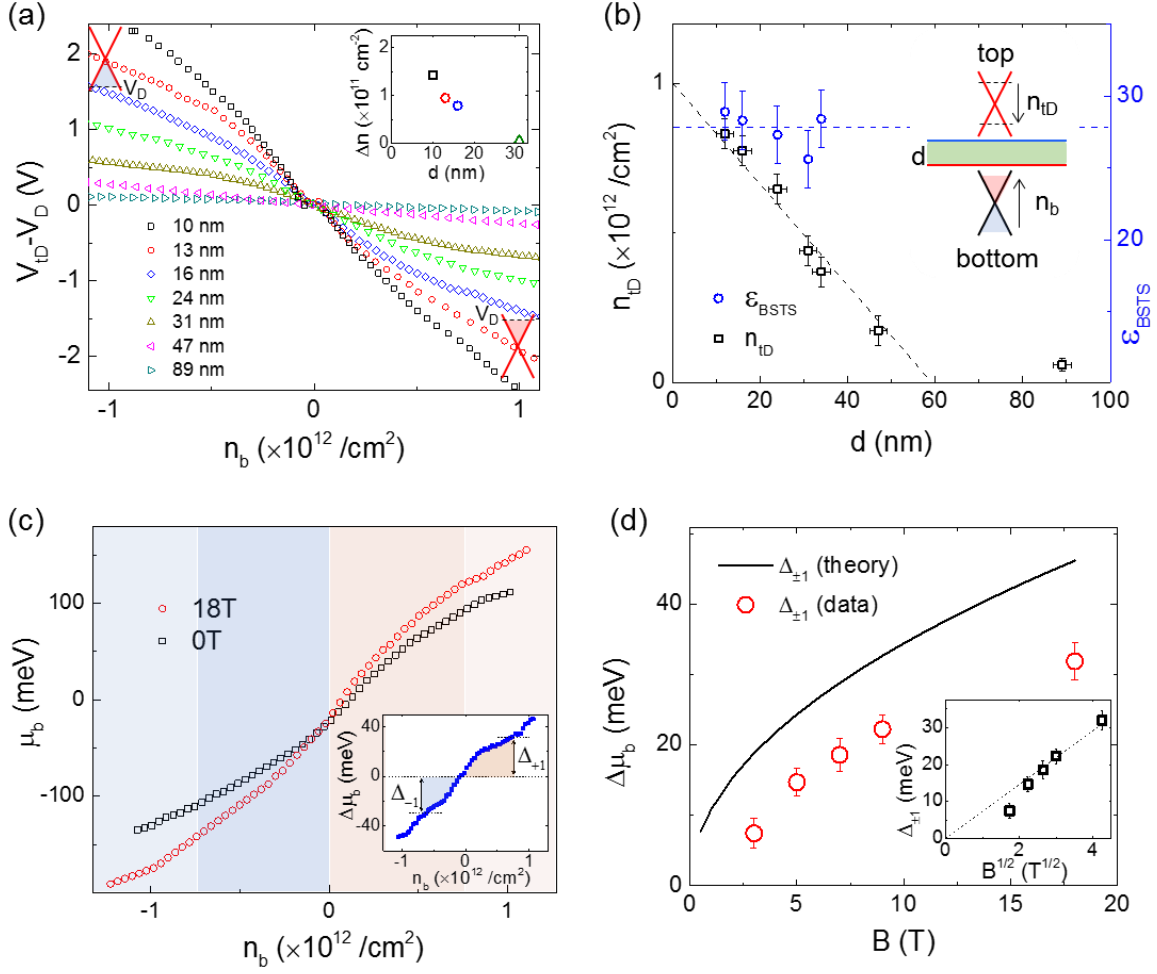

FIG. 3. (a) Plot of $V_{tD}-V_D$ as a function of $n_b$ for different thickness BSTS at 18 T. Inset in (a) plots the splitting of $N_t$ and $N_b$ at overall CNP in total charge density ($\Delta n$) as a function of flake thickness d. (b) Plot of $n_{tD}$ (as the $n_b$ is tuned from bottom Dirac point to $1\times10^{12}$ cm$^{-2}$) and $\varepsilon_{BSTS}$ as a function of d. The black and blue dashed lines in (b) are the fittings of $n_{tD}$ and $\varepsilon_{BSTS}$, respectively, with d. Inset in (b) is a schematic of the top and bottom Dirac surface states. (c) Plot of $\mu_b$ versus $n_b$ for the 16 nm BSTS device at magnetic field of 0T and 18T. The color highlights in (c) display the LL indices of bottom surface $N_b$ from -2 (leftmost) to +2 (rightmost) formed at 18 T. Inset in (c) is the $\Delta\mu_b = \mu_b(18T)-\mu_b(0T)$ versus $n_b$. (d) Plot of $\Delta\mu_b$ for $N_b=\pm1$ versus magnetic field. Inset in (d) is the $\Delta_{\pm1}$ versus $B^{1/2}$.